\documentclass[twocolumn,appendixfloats,tighten]{aastex61}

\usepackage{amssymb}
\usepackage{amsmath}
\usepackage{lipsum}
\usepackage{natbib}
 \usepackage{booktabs}

\newcommand{\mstar}{M_{\star}}
\newcommand{\mstart}{$M_{\star}$ }
\newcommand{\msun}{{\rm M_{\sun}}}
\newcommand{\FIREurl}{\url{http://fire.northwestern.edu}}

\received{July 12, 2017}
\revised{September 25, 2016}
\accepted{September 27, 2017}
\submitjournal{ApJ Letters}

\shorttitle{Stacked SFR profiles and bursty star formation}
\shortauthors{Orr et al.}

\begin{document}

\title{Stacked star formation rate profiles of bursty galaxies exhibit `coherent' star formation}

\correspondingauthor{Matthew Orr}
\email{meorr@caltech.edu}

\author{Matthew E. Orr}
\affiliation{TAPIR, Mailcode 350-17, California Institute of Technology, Pasadena, CA 91125, USA}

\author{Christopher C. Hayward}
\affiliation{Center for Computational Astrophysics, Flatiron Institute, 162 Fifth Avenue, New York, NY 10010, USA}
\affiliation{Harvard-Smithsonian Center for Astrophysics, 60 Garden Street, Cambridge, MA 02138, USA}

\author{Erica J. Nelson}
\affiliation{Max-Planck-Institut f\"ur Extraterrestriche Physik, Giessenbachstrasse, D-85748 Garching, Germany}

\author{Philip F. Hopkins}
\affiliation{TAPIR, Mailcode 350-17, California Institute of Technology, Pasadena, CA 91125, USA}

\author{Claude-Andr\'e Faucher-Gigu\`{e}re}
\affiliation{Department of Physics and Astronomy and CIERA, Northwestern University, 2145 Sheridan Road, Evanston, IL 60208, USA}

\author{Du\v{s}an Kere\v{s}}
\affiliation{Department of Physics, Center for Astrophysics and Space Science, University of California at San Diego, 9500 Gilman Drive, La Jolla, CA 92093, USA}

\author{T. K. Chan}
\affiliation{Department of Physics, Center for Astrophysics and Space Science, University of California at San Diego, 9500 Gilman Drive, La Jolla, CA 92093, USA}

\author{Denise M. Schmitz}
\affiliation{TAPIR, Mailcode 350-17, California Institute of Technology, Pasadena, CA 91125, USA}

\author{Tim B. Miller}
\affiliation{Department of Astronomy, Yale University, 52 Hillhouse Avenue, New Haven, CT, 06511, USA}



\begin{abstract}

In a recent work based on 3200 stacked H$\alpha$ maps of galaxies at $z \sim 1$, Nelson et al.~find evidence for `coherent star formation': the stacked SFR profiles of galaxies above (below) the `star formation main sequence' (MS) are above (below) that of galaxies on the MS at all radii. One might interpret this result as inconsistent with highly bursty star formation and evidence that galaxies evolve smoothly along the MS rather than crossing it many times. We analyze six simulated galaxies at $z\sim1$ from the Feedback in Realistic Environments (FIRE) project in a manner analogous to the observations to test whether the above interpretations are correct.  The trends in stacked SFR profiles are qualitatively consistent with those observed. However, SFR profiles of individual galaxies are much more complex than the stacked profiles: the former can be flat or even peak at large radii because of the highly clustered nature of star formation in the simulations. Moreover, the SFR profiles of individual galaxies above (below) the MS are not systematically above (below) those of MS galaxies at all radii. We conclude that the time-averaged coherent star formation evident stacks of observed galaxies is consistent with highly bursty, clumpy star formation of individual galaxies and is not evidence that galaxies evolve smoothly along the MS.

\end{abstract}




\section{Introduction} \label{sec:intro}

Given that star formation is one of the fundamental processes driving galaxy formation, it is crucial to understand what governs star formation, both on local and galactic scales. One of the open questions regarding star formation on galactic scales is whether it is coherent in space and/or time because of, e.g., gas accretion or environmental effects or highly stochastic because of, e.g., violent stellar feedback. The relatively tight correlation found between the star formation rate (SFR) and stellar mass ($\mstar$) of actively star-forming galaxies at a range of redshifts \citep{Brinchmann2004, Noeske2007, Peng2010, Wuyts2011}, commonly referred to as the `star formation main sequence' (MS), is sometimes taken as evidence of the former.  In particular, some authors argue that galaxies evolve smoothly along the sequence (rather than cross it), as is typically the case in large-volume cosmological simulations (such as those of the \emph{Illustris} and {\sc EAGLE} projects; \citealt{Vogelsberger2014,Schaye2015}) that rely on sub-grid ISM models. In such simulations, galaxies maintain their positions relative to the locus of the MS for $\gg 100$-Myr timescales \citep{Sparre2015illustris, Schaye2015}. However, high-resolution cosmological ``zoom-in" simulations that include explicit multi-channel stellar feedback suggest that star formation is very bursty in some regimes (due to the clustered nature of star formation, violent stellar feedback, galactic fountains, and stochastic gas inflow), including at high redshift. This burstiness causes galaxy-scale star formation to be a chaotic process in which galaxies cross the MS many times rather than evolve smoothly along it \citep{Hopkins2014fire,Muratov2015, Sparre2017, FG2017}.

Recent works \citep[e.g.,][]{Nelson2016b, Delgado2016} have investigated the average radial SFR profile of galaxies at a given mass and redshift by stacking H$\alpha$ maps of hundreds to thousands of galaxies. In particular, this work is motivated by the work of \citet{Nelson2016b}, who, based on a stacking analysis of 3200 galaxies, found evidence for what they term `coherent star formation': at a given mass and redshift, galaxies above (below) the MS have stacked SFR profiles above (below) those of MS galaxies at all radii; in contrast, their stellar mass profiles are nearly identical. This might be interpreted as evidence for smooth evolution of galaxies along and parallel to the MS, with coherent elevation (suppression) of star formation at all radii for galaxies above (below) the MS.  In other words, galaxies above (below) the main sequence remain above (below) the main sequence for long periods of time. This scenario is seemingly inconsistent with very bursty star formation, i.e., SFR variations of an order of magnitude or more on timescales $\la 100$ Myr.

To determine whether highly bursty star formation is consistent with the observations of \citet{Nelson2016b}, we investigate the radial SFR, stellar mass, and specific SFR (sSFR) surface density profiles of simulated galaxies from the Feedback in Realistic Environments (FIRE) project\footnote{\FIREurl}. We analyze the simulated galaxies in a manner analogous to the observations to understand the differences amongst the profiles of galaxies that lie above, on, and below the MS, and we compare individual galaxy profiles with the stacked profiles. We show that despite the star formation in the FIRE galaxies being highly bursty at the redshifts of interest,\footnote{In the FIRE simulations, galaxies with $\mstar \ga 10^{10}~\msun$ exhibit highly bursty star formation at high redshift and transition to steady star formation at $z \la 1$; lower-mass galaxies always exhibit bursty star formation \citep{Sparre2017}.} which causes them to cross the MS many times rather than evolve parallel to it, the stacked profiles exhibit trends similar to those observed. Consequently, we conclude that the time-averaged coherent star formation evident stacks of observed galaxies is consistent with highly bursty, clumpy star formation of individual galaxies and is not (necessarily) evidence that galaxies evolve smoothly along the MS.

\begin{figure*}
	\centering
	\includegraphics[width=0.94\textwidth]{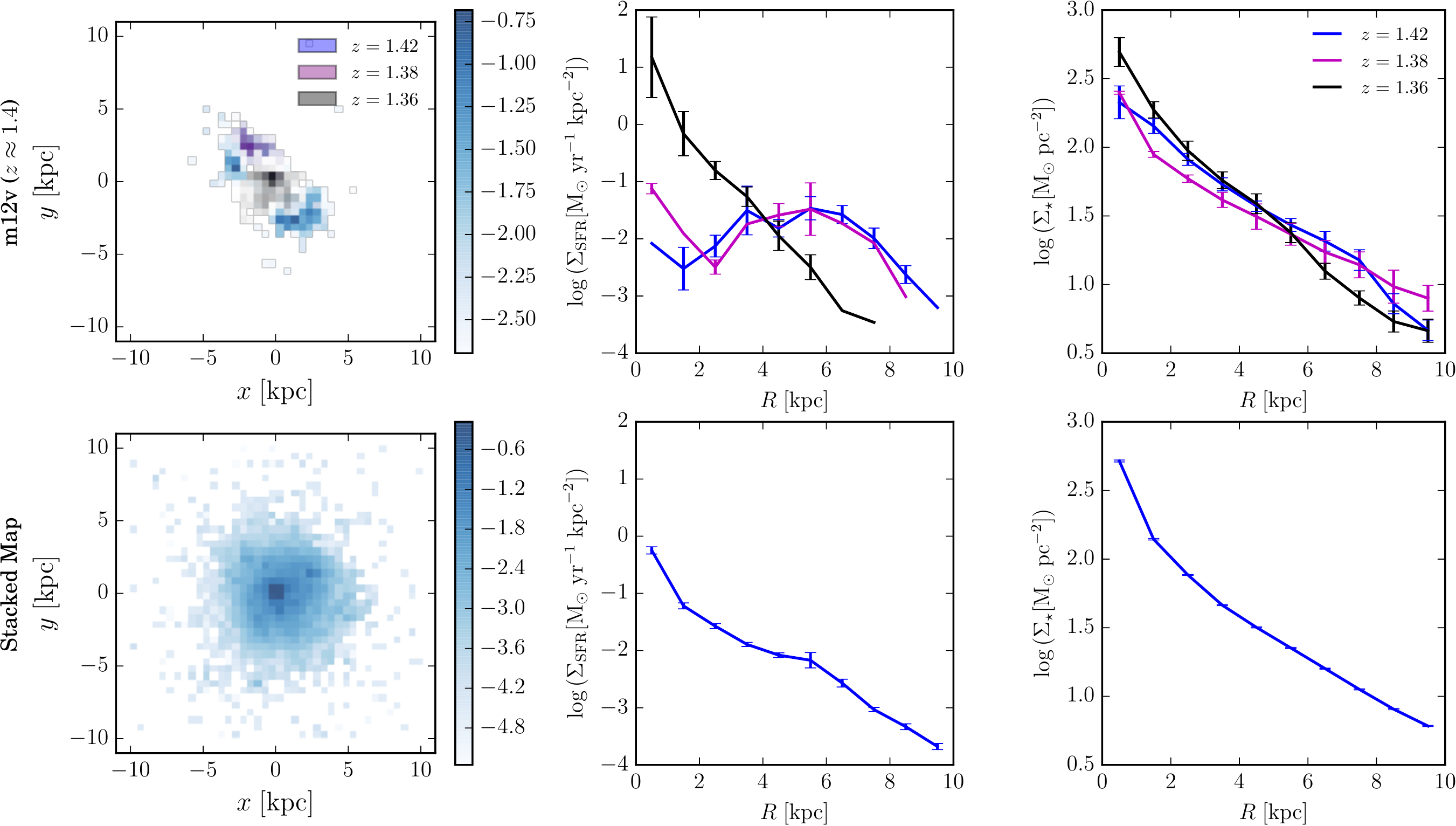}
	\caption{\emph{Top row}: three SFR surface density maps (\emph{left column}) and radial SFR (\emph{middle column}) and mass (\emph{right column}) profiles from different time slices of the `m12v' simulation around $z \approx 1.4$. \emph{Bottom row}: the results of stacking 205 snapshots in total from 3 different central galaxies with $\mstar \sim 10^{10}~\msun$ at redshifts $0.7 < z < 1.5$.  The individual galaxy's SFR maps reveal irregular, asymmetric, and highly time-variable SFR spatial distributions, and individual maps are often dominated by off-center star-forming clumps. In two cases, the radial SFR profiles have central peaks, but in one of those cases, the bulk of the star formation corresponds to the local maximum at $R\sim 5$~kpc.  The stacked map exhibits a clear central peak in SFR and has a monotonically radially decreasing SFR profile; it thus does not capture the diversity of SFR maps and profiles of single galaxy snapshots. In contrast, the individual galaxies' stellar mass profiles are all similar to the stacked stellar mass profile.}
	\label{maps}
\end{figure*}

\begin{figure}
	\centering
	\includegraphics[width=0.42\textwidth]{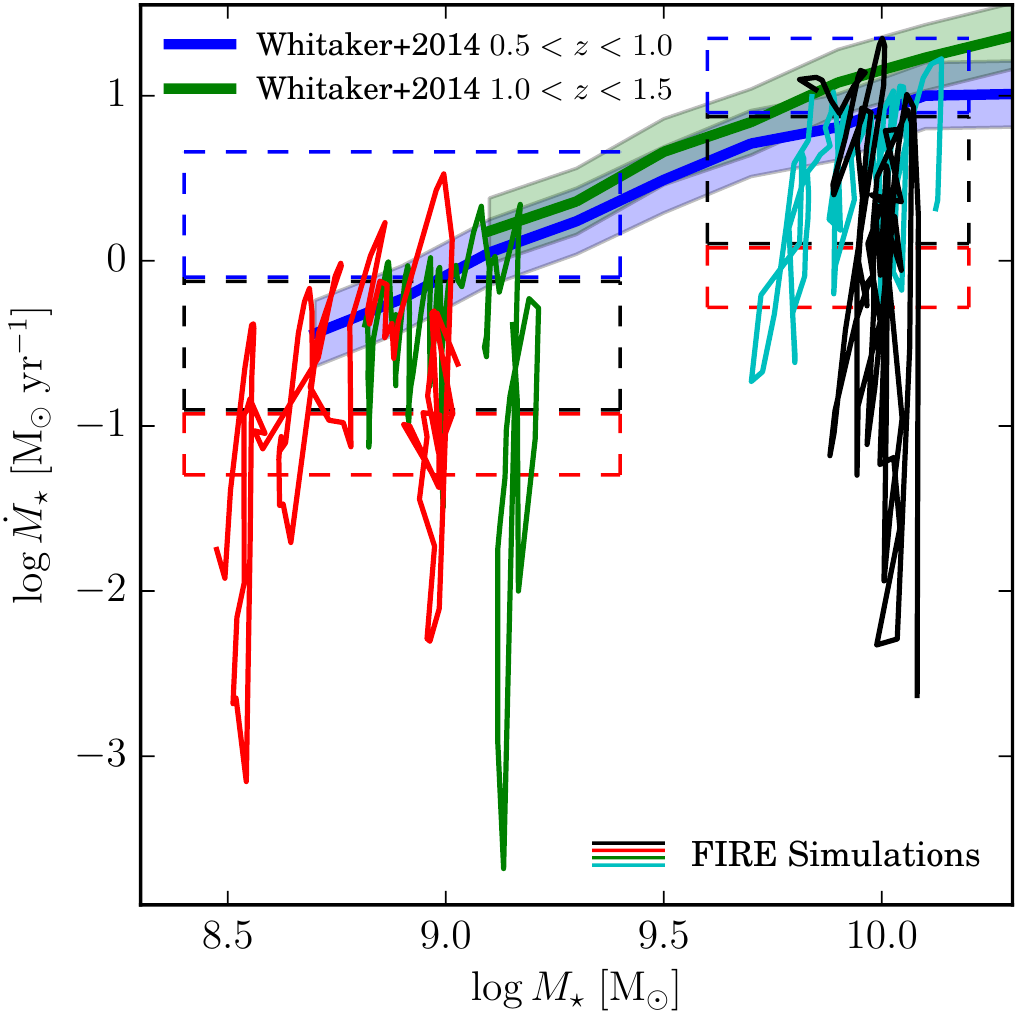}
	\caption{Tracks of two individual galaxy runs from each M$_\star$ bin (individually colored) in the SFR-$M_\star$ plane for $0.7 < z < 1.5$.  $\dot M_\star$ is the 100 Myr-averaged SFR within the central $20$ kpc of each main halo; $M_\star$ is calculated within the same aperture.  Solid blue (green) lines indicate the `star formation main sequence' (MS) in the redshift interval $0.5 < z < 1.0$ ($1.0 < z < 1.5$) found by \citet{Whitaker2014}; the shaded regions represent the intrinsic scatter of 0.2 dex found by \citet{Speagle2014}. The dashed colored boxes indicate the cuts used in this work to classify galaxies as above (blue), on (black), or below (red) the MS.  At these redshifts, FIRE galaxies have rapidly changing SFRs and do not evolve parallel to the MS but rather cross it many times.}
	\label{SFRMS}
\end{figure}

\begin{figure}
	\centering
	\includegraphics[width=0.42\textwidth]{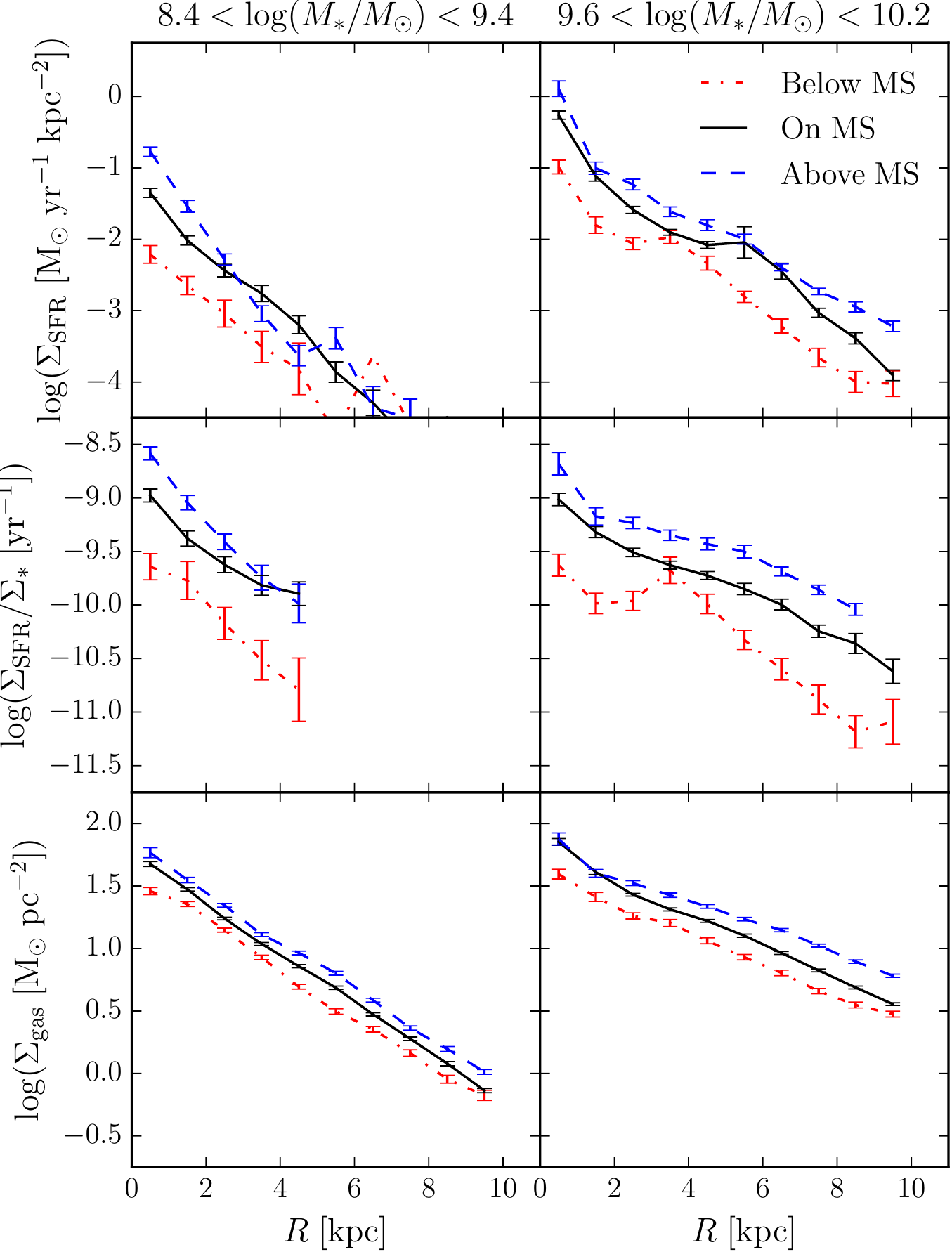}
	\caption{Stacked SFR (\emph{top row}), sSFR (\emph{middle row}), and neutral gas (\emph{bottom row}) surface density profiles (binned into 1~kpc annuli) for two stellar mass bins, $8.4 < \log (M_\star/{\rm M_\odot}) < 9.4$ (\emph{left column}) and $9.6 < \log (M_\star/{\rm M_\odot}) < 10.2$  (\emph{right column}), for $0.7 < z < 1.5$.  Prior to stacking, in each mass bin, the galaxies have been separated according to their position relative to the MS: above (\emph{blue dashed line}), on (\emph{black solid}), or below (\emph{red dash-dotted}).  The SFR, sSFR, and $\Sigma_{\rm gas}$ profiles generally decrease monotonically with radius.
	Moreover, the stacked profiles of galaxies above (below) the MS are above (below) those of MS galaxies at nearly all radii.}
	\label{sSFRprofiles}
\end{figure}

\begin{figure}
	\centering
	\includegraphics[width=0.42\textwidth]{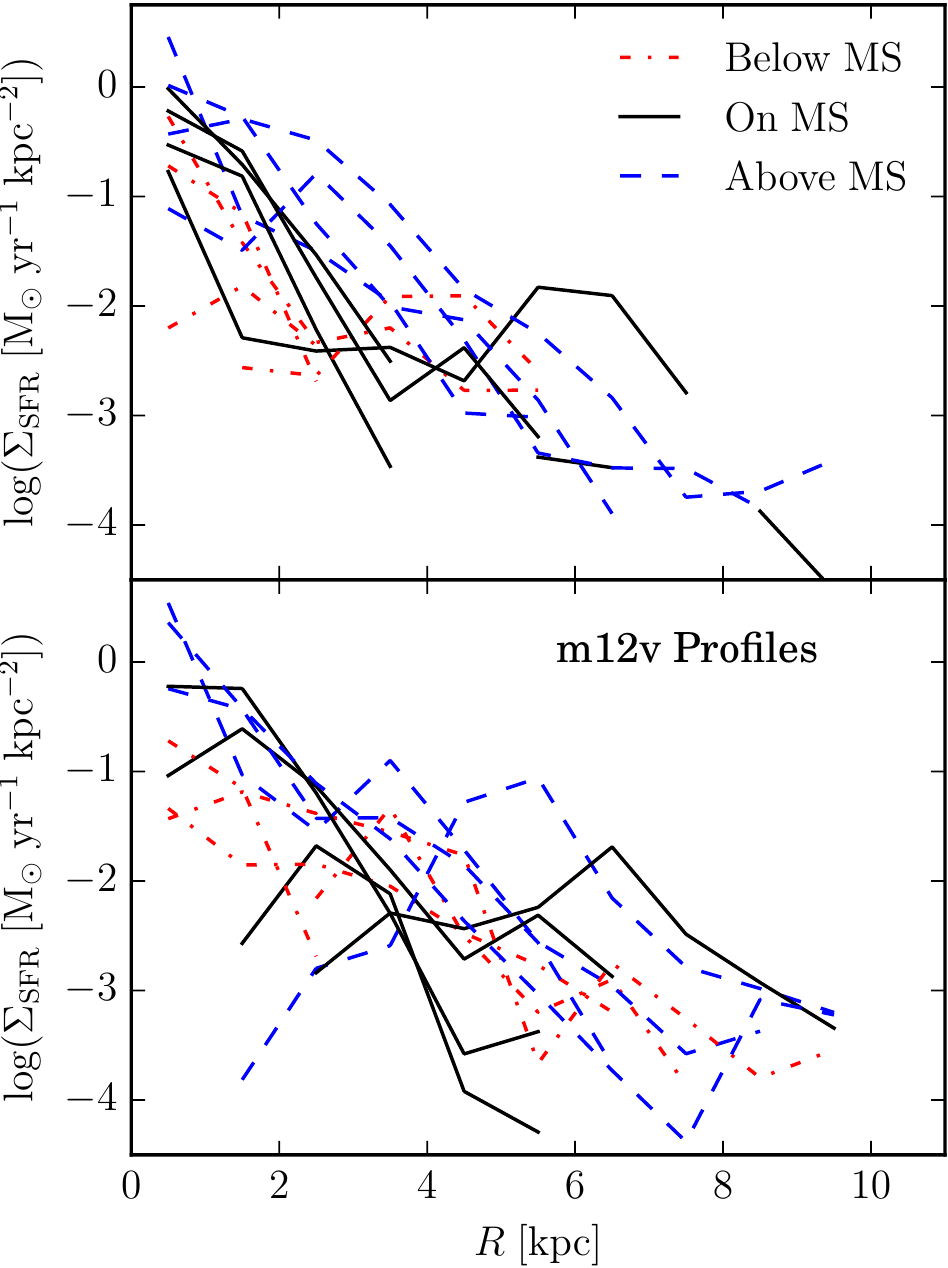}
	\caption{\emph{Top}: SFR surface density profiles of randomly selected individual snapshots (at $0.7 < z < 1.5$) with $9.6 < \log (M_\star/{\rm M_\odot}) < 10.2$, including those above (\emph{blue dashed}), on (\emph{black solid}), and below (\emph{red dash-dotted}) the MS (four of each type).  Error bars have been omitted for clarity.  The (10~Myr-averaged) SFR profiles of galaxies above (below) the MS are not systematically above (below) those of MS galaxies. Moreover, in some cases, the profiles peak at large radii. \emph{Bottom}: Consistent results are seen for randomly selected snapshots of a single galaxy run, `m12v', in the same redshift and mass bins.  Stacking reflects the fact that star formation in the simulated galaxies is coherent in a time-averaged sense even though individual galaxies evolve in a bursty manner and not parallel to the MS.}
	\label{gals}
\end{figure}

\section[]{Methods}\label{meth}

We investigate the radial star formation profiles of a selection of the FIRE-1 galaxy simulations originally presented in \citet{Hopkins2014fire} and \citet{Chan2015}, which were run using {\sc gizmo} \citep{Hopkins2015gizmo} in its pressure-energy smoothed particle hydrodynamics (P-SPH) mode \citep{Hopkins2013a}.  The physics, source code, and all numerical parameters are identical to those in all other FIRE-1 simulations. The simulations incorporate cooling from $10-10^{10}\,$K, including atomic, molecular, and metal-line cooling processes and accounting for photo-heating by a UV background \citep{FaucherGiguere2009}, in addition to self-shielding. Stars form only in dense ($n \ga 50$ cm$^{-3}$), self-gravitating, self-shielding, molecular gas. Multi-channel stellar feedback from supernovae, radiation pressure from massive stars, stellar winds, and photo-ionization/heating is treated explicitly based on the outputs of the {\sc starburst99} \citep{Leitherer1999} stellar evolution models, assuming a \citet{Kroupa2001} IMF.  The stellar and gas masses and stellar half-mass radii of the simulations analyzed here at $z \approx 1$ are presented in Table~\ref{simprops}.
\newcommand{\ra}[1]{\renewcommand{\arraystretch}{#1}}
\ra{0.9}
\begin{table}[]
\centering
\caption{Simulation Properties at $z=1$}
\label{simprops}
\begin{tabular}{@{}lccc@{}}
\hline
Name    	& $M_\star$ 		& $M_{\rm gas}$ 		& $R_{\rm half}$ 	\\
		& $(10^9 \rm \, M_\odot)$	& $(\rm 10^9 \, M_\odot)$	& $({\rm kpc})$		\\ \hline
m11h383$^\dagger$ & 1.1	& 2.3		& 2.9 \\
m11         & 0.81	& 2.3 	& 5.4 \\
m11v       & 1.7 		& 1.2 	& 5.9 \\
m12q       & 13		& 2.3 	& 3.3 \\
m12i        & 12		& 6.8 	& 5.4 \\
m12v       & 13		& 2.7		& 6.4 \\ \hline
\multicolumn{4}{l}{(1) Name: simulation designation.} \\
\multicolumn{4}{l}{(2-3) $M_\star$,$M_{\rm gas}$: Stellar \& gas masses in maps.}\\
\multicolumn{4}{l}{(4) $R_{\rm half}$: Stellar half-mass radius.} \\
\multicolumn{4}{l}{$^\dagger$ Except for m11h383 \citep{Chan2015},} \\
\multicolumn{4}{l}{all simulations are from \citet{Hopkins2014fire}.
}
\end{tabular}
\end{table}

To probe the radial SFR profiles in the simulations, we use spatially resolved face-on projected maps of SFR and stellar mass surface density from simulated galaxies spanning redshifts  $z = 0.7-1.5$ produced by \citet{Orr2017a}.  To compare the snapshots with the observations of \citet{Nelson2016b}, we use maps with 1~kpc$^2$ pixels centered on the centers of the stellar mass distributions in the snapshots. 10 Myr-averaged SFR maps are computed by summing the stellar mass in young ($< 10$ Myr) star particles in each pixel, correcting for the mass lost due to stellar evolution effects using {\sc starburst99} \citep{Leitherer1999}, and dividing by 10 Myr. This time interval approximately corresponds to the timescale traced by recombination lines such as $\rm H\alpha$ \citep{Kennicutt2012}.
\section{Results}\label{results}

Fig.~\ref{maps} shows several individual SFR surface density maps and the result of stacking many maps from the same \mstart bin, in addition to their respective radially averaged SFR and stellar mass surface density profiles.  The top row shows three of the 10 Myr-averaged SFR maps of the `m12v' central galaxy from \citet{Hopkins2014fire} at $z\approx 1.4$, with $\mstar \sim 10^{10} ~\msun$, with 1~kpc pixel sizes, and their associated radially averaged SFR and stellar mass surface density profiles.  The SFR profiles of the galaxy vary considerably from $z = 1.42-1.36$ and are not always centrally peaked. The bottom row shows the result of stacking the SFR maps of 205 snapshots in total, from 3 distinct galaxies ($\sim 70$ from each, with $\Delta z = 0.01$ spacing\footnote{For $0.7 < z < 1.5$, the snapshot spacing of $\Delta z = 0.01$ corresponds to time spacing of $25-56$ Myr.}) in the $9.6 < \log (M_\star/{\rm M_\odot}) < 10.2$ stellar mass bin. For all radial profiles shown in this work, we compute the error bars by bootstrap resampling the pixels in the annuli following \citet{Nelson2016b}.  In the stacked map, we see a much smoother and more azimuthally symmetric spatial distribution, and the corresponding radially averaged profiles are smoother, monotonically decreasing functions of radius.  Here, by averaging hundreds of snapshots of galaxies of similar mass, we recover the fact that the simulated galaxies have higher gas densities in their centers, and thus form more stars there on average. However, the SFR profiles of individual galaxies at a given time can differ dramatically from the stacked profile.  In contrast, the stacked stellar mass profile is fairly representative of the individual profiles.

Fig.~\ref{SFRMS} shows tracks of four simulated galaxies in the SFR-$M_\star$ plane for $0.7 < z < 1.5$; the 100 Myr-averaged SFR is employed. The observed MS \citep{Whitaker2014} and scatter \citep{Speagle2014}
in two redshift bins intersecting this interval are shown.
In this redshift and mass range, the individual simulated galaxies experience significant (sometimes order of magnitude or more), rapid (timescales $\la 100$ Myr) SFR variations (see \citealt{Sparre2017} for
a detailed study) and clearly do not evolve parallel to the MS.\footnote{The
galaxies' stellar masses do not increase monotonically in time because the stellar mass is computed within a radius of 20 kpc from the halo center; when satellites leave the aperture, the total stellar mass can decrease.}

Following \citet{Nelson2016b}, we label individual snapshots as being above, below, or on an MS determined by the distribution of the galaxy-integrated 100~Myr-averaged SFRs in a given \mstart bin (because \citealt{Nelson2016b} classify galaxies relative to the MS according to their UV+IR SFRs, and for actively star-forming galaxies, this indicator traces the SFR over the past $\sim 100$~Myr; \citealt{Kennicutt2012,Hayward2014}).
For a given \mstart bin, we rank the galaxies by SFR and consider the median value to be the locus of the MS.  We then employ the same SFR cuts as \citet{Nelson2016b}, defining galaxy snapshots within $\pm 0.4$~dex of the median SFR to be on the MS and those $+0.4-1.2$~dex and $-(0.4-0.8)$~dex away to be above and below the MS, respectively.  These cuts are represented by the dashed colored boxes in Fig.~\ref{SFRMS}.

We then stack individual galaxy maps according to their position with respect to the MS in two bins of $\mstar$, producing average SFR, specific SFR, and neutral gas (atomic + molecular) surface density profiles, which are presented in Fig.~\ref{sSFRprofiles}.  We see that for both \mstart bins, the stacked SFR profile of galaxies above (below) the MS is above (below) the SFR profile of galaxies on the MS at nearly all radii, i.e., star formation appears to be coherently enhanced (suppressed) at nearly all radii in galaxies above (below) the MS. Moreover, the stacked SFR profiles exhibit a similar approximately exponential shape, peaking in the center and declining with radius, regardless of position with respect to the MS.  The stacked stellar mass surface density profiles are nearly identical in each \mstart bin for all classes of galaxies, so we do not show them. 
The sSFR surface density profiles in Fig.~\ref{sSFRprofiles} also exhibit a clear separation by class. The neutral gas surface density profiles also vary systematically across the MS, but the difference between above- and below-MS galaxies is considerably less than for the sSFR profiles.
We note that the results of Fig.~\ref{sSFRprofiles} do not qualitatively change when the maps are re-normalized by their half-mass radii before stacking, indicating that these results are somewhat robust to evolution within the redshift interval and to the particular manner of stacking.  We conclude that coherent star formation is apparent in the stacked SFR profiles despite the underlying galaxies exhibiting very bursty star formation and often having their total SFRs dominated by individual off-center clumps.

One apparent tension between the observations and simulations is that in the simulations, the sSFR profiles are generally centrally peaked, whereas the stacked H$\alpha$ equivalent width profiles of observed galaxies are flat \citep{Nelson2016b, Tacchella2017}. This tension may be partially due to dust attenuation (see \citealt{Nelson2016a}), especially for above-MS galaxies, which may have significant central dust-obscured star formation \citep{Wuyts2011,Hemmati2015}. However, it is not clear that correcting for dust would resolve the discrepancy, especially for lower-mass galaxies, and this issue deserves further attention. Another possible reason is that in low-mass galaxies, our centering on the stellar center of mass likely differs from the centering in the observations (based on light), which is likely affected by lumpy/irregular morphologies and local variations in mass-to-light ratio; this effect may cause the observed stacked profiles to be artificially flat. 

To connect the stacked SFR profiles with those of individual galaxies at a given time, we examine a randomly chosen sub-sample of the individual radial SFR profiles in the $9.6 < \log (M_\star/{\rm M_\odot}) < 10.2$ stellar mass bin in the top panel of Fig.~\ref{gals}.\footnote{Not all of the profiles reach the centers of the galaxies because some have identically zero SFR at their centers.} Although galaxies classified as above the MS have greater 100~Myr-averaged SFR values than those on or below the MS, there is significant crossing of the (10~Myr-averaged) SFR profiles at modest galactocentric radii, i.e., the SFR profiles of individual galaxies above (below) the MS are typically not systematically above (below) those of MS galaxies. There does not appear to be significant differences in the \emph{forms} of SFR profiles amongst these classes of galaxies in the FIRE simulations; only their relative normalization differs, a feature that \citet{Nelson2016b} describe as `coherent star formation'. By selecting galaxies above (or below) the MS, we tend to select galaxies just as they are forming many stars in a burst (are in a relatively quiescent period). The bottom panel of Fig.~\ref{gals}, which shows radial SFR profiles of a single galaxy \citep[`m12v' from][]{Hopkins2014fire} at different randomly drawn times within the redshift interval  $0.7 < z < 1.5$ (four each above, on and below the MS), reinforces this conclusion.
The galaxy's SFR profile varies rapidly with time, and there is no clear dependence on the total SFR (i.e., position relative to the MS).

\section{Summary and discussion}\label{sum}

We have analyzed the individual and stacked SFR maps and profiles of a sample of simulated galaxies from the FIRE project in a manner analogous to the observational analysis of \citet{Nelson2016b}. Despite the FIRE galaxies exhibiting large variations in SFR on $\sim 10-100$ Myr timescales and often having their SFRs concentrated in a few massive off-center clumps, their stacked SFR profiles exhibit spatial coherent star formation in a time- and azimuthally averaged sense.  Moreover, individual SFR profiles in the FIRE simulations often look nothing like the stacked profiles.  A similar effect has been seen in observations: Fig. 4 of \citet{Nelson2016b}, for example, shows that the individual H$\alpha$ maps combined into stacks exhibit a variety of different morphologies.  Moreover, the stacked SFR profiles of simulated galaxies above (below) the MS are above (below) those of the MS galaxies at all radii.  This is consistent with the observations of \citet{Nelson2016b}, indicating that in simulations with resolved ISM and bursty stellar feedback, star formation can still be coherent in a time-averaged sense. We stress that in the mass and redshift ranges considered, the FIRE galaxies cross the MS many times throughout their evolution due to their highly bursty star formation histories; thus, one should not interpret the appearance of coherent star formation in stacked SFR profiles as evidence that galaxies maintain their positions relative to the MS for long periods of time.

There are two main lessons from this analysis. First, although stacking recovers the time-averaged spatial coherence of star formation in the simulations, it hides the chaotic, incoherent nature of star formation on kiloparsec scales. In the simulations, the SFR is on average higher in the centers of galaxies, owing to galaxies typically having centrally peaked gas surface density profiles; the stacked profiles recover this average behavior. However, the bursty nature of star formation in the FIRE galaxies, in which the SFR at a given time can be dominated by a few short-lived \citep[$\sim 20$ Myr;][]{Oklopcic2017,Sparre2017, FG2017} massive clumps of star formation at various galactocentric radii, is obscured by the stacking procedure. We indeed find that the stacking analysis makes stochastic enhancements in the SFR from massive clumps, which are often located significantly off-center, indistinguishable from global enhancements in the SFR across the disc; this possibility was noted in \citet{Nelson2016b}. 
 
Second, the simulations discussed here provide insight into what causes galaxies to be above or below the MS.  In Fig.~\ref{gals}, we see that galaxies selected to be above the MS have preferentially recently formed several massive clumps of stars; this is true whether the SFR is averaged over 10 or 100 Myr (i.e., whether H$\alpha$- or UV+IR-based SFRs are used). Conversely, galaxies below the MS are unlikely to have formed many massive clumps within the past $\sim 100$ Myr and rather are likely to be in a low-sSFR period, which can last for a few 100s of Myr in the simulations \citep{Muratov2015, Sparre2017}; if they have formed a few clumps, the associated SFRs are not as high as in the above-MS galaxies. Moreover, on average, the below-MS galaxies tend to have lower SFRs at all radii than above-MS galaxies (but this is not true of the individual profiles) because these galaxies have, on average, lower gas surface densities (Fig.~\ref{sSFRprofiles}) than the above-MS galaxies (owing to stochasticity in gas accretion from both the IGM and galactic fountains and/or recent strong outflows driven by stellar feedback; \citealt{Muratov2015, AA2016, Hayward2017}). However, in the simulations, these differences are stochastic rather than long-lived, as evident from the bottom panel of Fig.~\ref{gals}, and the FIRE galaxies can cross the MS multiple times within 100 Myr \citep[Fig.~\ref{SFRMS}; see also][]{Sparre2017}.  Note, however, that the galaxies considered here are of relatively `low' stellar mass (M$_\star \lesssim 10^{10}$ M$_\odot$), and more massive simulated galaxies tend to exhibit less bursty star formation and smoother mass, metallicity, and SFR profiles, especially at low redshift \citep{Sparre2017, Ma2017}.

We find that very bursty star formation is consistent with spatially coherent star formation in stacked images.  We thus caution against interpreting such time-averaged coherent star formation as evidence that galaxies maintain their positions relative to the MS owing to, e.g., systematic differences in gas accretion rates and thus gas fractions.
A crucial next step is to place observational constraints on the timescale over which galaxies oscillate across the main sequence, perhaps via measurement of SFR tracers that probe different timescales \citep[e.g.,][]{Guo2016}. Although our analysis does not rule out the possibility that galaxies maintain their positions relative to the MS for long periods of time, it demonstrates that simulations in which this is not the case yield stacked SFR profiles consistent with those observed, including spatially coherent star formation in a time-averaged sense. More generally, our analysis demonstrates that simulations are a valuable tool that can help understand behaviors of individual galaxies that may be masked in stacked observations.

\acknowledgments

MEO is grateful for the encouragement of his late father, SRO, in studying astrophysics, and is supported by the National Science Foundation Graduate Research Fellowship under Grant No. 1144469.  We are grateful to the anonymous referee for providing us with constructive comments and suggestions.  The Flatiron Institute is supported by the Simons Foundation.  Support for PFH was provided by an Alfred P. Sloan Research Fellowship, NASA ATP Grant NNX14AH35G, and NSF Collaborative Research Grant \#1411920 and CAREER grant \#1455342.  CAFG was supported by NSF through grants AST-1412836 and AST-1517491, by NASA through grant NNX15AB22G, and by STScI through grant HST-AR-14562.001. DK acknowledges support from NSF grant AST-1412153 and the Cottrell Scholar Award from the Research Corporation for Science Advancement.  DMS is supported by the National Science Foundation Graduate Research Fellowship under Grant No. 2015192719.

\end{document}